\documentclass[12pt]{article}
\usepackage{epsfig}
\usepackage{graphicx}
\usepackage{amsmath}
\usepackage{amsfonts}
\usepackage{lscape}
\usepackage{epsfig}
\usepackage{natbib}

\def\nn{\nonumber}
\def\ni{\noindent}
\def\wh {\widehat}

\newtheorem{theorem}{Theorem}
\begin{document}

\clearpage\pagebreak
\baselineskip=24pt
\thispagestyle{empty}

\begin{center}
\Large{A mixture distribution approach for assessing genetic impact from twin study}        
\end{center}

\vskip 1.5cm

\begin{center}
Zonghui Hu$^{1*}$, Pengfei Li$^{2*}$, Dean Follmann$^{1}$, Jing Qin$^{1}$ \\
$^1$ Biostatistics Research Branch, National Institute of Allergy and Infectious Diseases, National Institutes of Health,  \\ Rockville, MD 20852, USA \\
$^2$ Department of Statistics and Actuarial Science, University of Waterloo, Waterloo, ON, Canada \\
$*$ Authors contribute equally
\end{center}

\begin{center}
ABSTRACT 
\end{center}

This work was motivated by a twin study with the goal of assessing the genetic control of immune traits.  We propose a mixture bivariate distribution to model twin data where 
 the underlying order within a pair is unclear.  Though estimation from mixture distribution is usually subject to low convergence rate, the combined likelihood, which is constructed over monozygotic and dizygotic twins combined, reaches root-n consistency and allows effective statistical inference on the genetic impact.  The method is  applicable to general unordered pairs. 

\begin{description}
\item[Keywords:]
 Correlation coefficient;  Maximum likelihood estimation; Mixture distribution; Unordered pair
\end{description}

\newpage
\setcounter{page}{1}
\section{Introduction} \label{sec:intro}

It is difficult to evaluate the genetic impact on a biological feature, such as the characteristics of human behavior or the strength of immune system,  and separate them from environmental impact.  This is especially true when there is little knowledge about what genetic factors contribute to a feature, aka trait,  and no genotypic data are actually available. Twin studies provide  a good setting for tackling the problem by exploiting the unique degree of genetic and environmental sharing among  two types of twins: the monozygotic (MZ) twins who are genetically identical, and the dizygotic (DZ) twins who share the same genetic similarity as ordinary siblings.  Since twins typically share a common environment,   it is possible to evaluate the extent to which genetic factors influence a trait by comparing MZ and DZ twins.   

This work was motivated by a study on immune traits.   The immune system functions through different immune cells.  Existing studies have shown that frequencies, or the proportionate representation, of major immune cell subsets constitute the canonical traits of immune system \citep{Ahmadi2001}.  To evaluate the genetic control of immune traits,  a study was conducted which included $324$ twins: $96$ MZ twins and $288$ DZ twins \citep{Mario2015}.  The goal was to assess the genetic impact on each immune trait, which helped to identify the genetically determined traits.

 For a specific trait, let $(Y^{*}_{i1}, Y^{*}_{i2})$ stand for the paired data in the $i$-th twin.  Let $\rho_{M}$ and $\rho_{D}$ stand for the correlation coefficient, ${\rm cor}(Y^{*}_{i1}, Y^{*}_{i2})$, in MZ and DZ twins, respectively.  The collective genetic impact is assessed by 
$\delta = \rho_{M} - \rho_{D}$: the larger $\delta$, the greater genetic impact on the trait. This is the well-known Falconer's formula, which assesses the genetic impact by effectively removing environmental influences \citep{Falconer1996}. 

However, the underlying order within a pair, which is determined by latent genetic factors, is unknown in twin data \citep{Ernst1996}. The observed are pairs $\{ (Y_{i1}, Y_{i2}),  i=1, \cdots, n \}$, where $(Y_{i1}, Y_{i2})$ can be either $(Y_{i1}^{*}, Y_{i2}^{*})$ or  $(Y_{i2}^{*}, Y_{i1}^{*})$, or known as the unordered pairs \citep{Hinkley1973, Davies1988}. Since MZ twins are genetically identical, this order is exchangeable in MZ twins. However, this order is not exchangeable in DZ twins. 
Though Pearson's and Spearman's are commonly used for assessing correlation, they require right labeling of measurements in a pair, with each measurement corresponding to one variable or one category. Changing the order in a pair can substantially alter the results.   Order does not matter to MZ twins but matters to DZ twins.  To handle the issue of missing order, statistical methods are usually based on ``symmetric'' metrics, such as the intraclass correlation coefficient \citep{sedgwick2013}.  These approaches attempt to avoid biased inference, but at the cost of removing the underlying order in paired data. 

In this work, we exploit the properties of twin data and directly target the missing order.  We model   twin data by a bivariate mixture distribution which effectively accounts for the missing order in each pair and  estimate through two types of likelihood functions: one is the separate likelihood built upon MZ and DZ twin data separately, and the other is the combined likelihood built upon MZ and DZ data combined. The different between the two is that the latter utilizes the property of equal variance shared by MZ and DZ twins.  Both maximum likelihood estimators are consistent.  Inheriting the limitation of mixture distribution, the separate likelihood estimator has slow convergence.  However, the combined likelihood estimator has improved efficiency, maintains root-n consistency in $\rho_{M}$ and $\rho_{D}$, and allows  for effective statistical inference on the collective genetic impact. 
 The bivariate mixture model and the likelihood functions are introduced in Section \ref{sec:mixture}.  
The asymptotic properties are studied in Section \ref{sec:combined}. Numerical studies and  application to the twin study on immune traits are presented in Section \ref{sec:simulation} and Section \ref{sec:application}.   Though methods and theories developed in this work arise from twin study, they apply to general unordered paired data when  assessment of correlation is the interest.

\section{Mixture bivariate distribution and likelihood function} \label{sec:mixture}
It is plausible to assume that $(Y^{*}_{1}, Y^{*}_{2})$ follows a bivariate normal distribution $\phi(y^{*}_{1}, y^{*}_{2};\mu_1,\mu_2,\rho,\sigma^2)$.   
 For MZ twins, $\mu_{1}=\mu_{2}$. For DZ twins,  $\mu_{1}$ and $\mu_{2}$ may or may not be equal: a trait with 
 $\mu_{1} = \mu_{2}$ is referred as a  homogeneous trait and one with $\mu_{1} \neq \mu_{2}$ a heterogeneous trait.

Since the underlying order is unknown, data are recorded in a random order.  In each twin, the observed are paired data $(Y_{i1}, Y_{i2})$, which has the same values as $(Y^{*}_{i1}, Y^{*}_{i2})$ but may be swapped; that is, $Y_{i1}=Y^{*}_{i2}, Y_{i2}=Y^{*}_{i1}$. They thus follow a mixture bivariate normal distribution 
\begin{eqnarray}
\hspace{-0.1in}
f(y_{1}, y_{2}; \mu_{1}, \mu_{2}, \rho, \sigma^{2}) =   (1- \alpha) \phi(y_{1}, y_{2}; \mu_{1}, \mu_{2}, \rho, \sigma^{2}) +  \alpha   \phi(y_{1}, y_{2}; \mu_{2}, \mu_{1}, \rho, \sigma^{2}), 
\label{eq:mixture}
\end{eqnarray} 

\ni where it is safe to fix $\alpha=0.5$ as labeling is random.  This mixture distribution applies to both MZ and DZ twins.  

Under the mixture distribution, both $Y_{1}$ and $Y_{2}$ follows the same mixture distribution $0.5 g(y; \mu_{1}, \sigma) + 0.5 g(y; \mu_{2}, \sigma)$ with $g$ the density of univariate normal distribution and 
\begin{eqnarray}
 E(Y_{1}) = E(Y_{2}) &=& 0.5(\mu_{1} + \mu_{2}), \quad E(Y_{1}Y_{2}) =  E(Y_{1}^{*}Y_{2}^{*}) 	=  \rho \sigma^{2} + \mu_{1} \mu_{2} \nn \\
 && E(Y_{1}^{2}) = E(Y_{2}^{2}) = \sigma^{2} + 0.5(\mu_{1}^{2} + \mu_{2}^{2}) \nn 
\end{eqnarray}

Pearson's correlation coefficient applied to $\{(Y_{i1}, Y_{i2}), i=1, \cdots, n\}$ is
\[
\frac{\sum_{i=1}^{n} (Y_{i1} - \bar{Y}_{.1})((Y_{i1} - \bar{Y}_{.2})} {\sqrt{\sum_{i=1}^{n} (Y_{i1} - \bar{Y}_{.1})^{2}}\sqrt{\sum_{i=1}^{n} (Y_{i2} - \bar{Y}_{.2})^{2}} }
\]

\ni where $ \bar{Y}_{.1}$ and $ \bar{Y}_{.2}$ stand for the means over the first and the second measurement in the unordered pairs, respectively. It converges to 
\begin{equation}
\frac{\rho \sigma^{2} -0.25(\mu_{1} - \mu_{2})^{2}} {\sigma^{2} + 0.25(\mu_{1} - \mu_{2})^{2}},
\label{eq:Pearson}
\end{equation}

\ni underestimating $\rho$ unless $\mu_{1}=\mu_{2}$: the more $\mu_{1}$ differs from $\mu_{2}$, the more bias.

We propose to estimate correlation based on the mixture distribution. For MZ twins,  $\mu_{1} = \mu_{2}=\mu_{M}$ and  (\ref{eq:mixture}) reduces to bivariate normal distribution $\phi(y_{1}, y_{2}; \mu_{M}, \mu_{M}, \rho_{M}, \sigma_{M}^{2})$.   For DZ twins, (\ref{eq:mixture}) can be more specifically written as $f(y_{1}, y_{2}; \mu_{D1}, \mu_{D2}, \rho_{D}, \sigma_{D}^{2})$.  We can estimate $\rho_{M}$ and $\rho_{D}$ from (\ref{eq:mixture}) over MZ and DZ data separately.

It is reasonable to assume the same variance in MZ and DZ twins, $\sigma_{M}^{2} = \sigma_{D}^{2}$ $=\sigma^{2}$.  Letting $\{ (Y_{i1}^{M}, Y_{i2}^{M}): i=1, \cdots, n_{M}\}$ stand for the MZ twin data and  $\{ (Y_{j1}^{D}, Y_{j2}^{D}): j=1, \cdots, n_{D}\}$ the DZ twin data, we construct the combined likelihood 
\begin{eqnarray}
l^{C}_n(\mu_{M}, \rho_{M},  \mu_{D1}, \mu_{D2},  \rho_{D}, \sigma^{2}) = 
l^{M}_n(\mu_{M},   \rho_{M}, \sigma^{2})  +   l^{D}_n(\mu_{D1}, \mu_{D2},   \rho_{D}, \sigma^{2} )
\label{eq:combined}
\end{eqnarray}
with
\[
l^{M}_n(\mu_{M},   \rho_{M}, \sigma^{2})
=\sum_{i=1}^{n_M}\log\{
\phi(Y_{i1}^{M}, Y_{i2}^{M};\mu_{M},\mu_{M}, \rho_{M}, \sigma^{2})
\}
\]
and 
\[
 l^{D}_n(\mu_{D1}, \mu_{D2},   \rho_{D}, \sigma^{2} ) =   \sum_{j=1}^{n_D}\log\{
\phi(Y_{j1}^{D}, Y_{j2}^{D};\mu_{D1}, \mu_{D2}, \rho_D, \sigma^{2})
+
\phi(Y_{j1}^{D}, Y_{j2}^{D};\mu_{D2}, \mu_{D1},\rho_D,\sigma^{2})\}. 
\]

\ni In the above,  $l^{M}_{n}$ is the log-likelihood over MZ data and $l^{D}_{n}$ 
the log-likelihood  over DZ data up to  a constant.    

We refer to (\ref{eq:mixture}) over MZ and DZ data separately as the separate mixture and (\ref{eq:combined}) the combined mixture.  

\section{Asymptotic properties} \label{sec:combined}

 \subsection{Separate mixture likelihood estimation}  
 
Estimation of (\ref{eq:mixture}) over MZ data is the estimation of  bivariate normal  $\phi(y_{1}, y_{2}; \mu_{M}, \mu_{M}, \rho_{M}, \sigma_{M}^{2})$.  The maximum likelihood estimator (MLE)  of $\rho_{M}$ is root-n consistent and equivalent to Pearson's correlation coefficient. The more complicated is the estimation of (\ref{eq:mixture}) over DZ data. Let $(\wh{\mu}_{D1}, \wh{\mu}_{D2},  \wh{\rho}_{D}, \wh{\sigma}^{2}_{D})$ stand for the MLE's of DZ related parameters.  If  $\mu_{D10} \neq \mu_{D20}$, the MLE's  are asymptotically normally distributed with root-$n$ consistency following the standard properties of MLE. 
If in fact $\mu_{D10}$ $=$ $\mu_{D20}$,  the separate likelihood estimation is subject to the limitation of mixture distribution, namely the slow convergence. Denote by $(\mu_{D0}, \rho_{D0}, \sigma_{D0}^{2} )$ the true parameter values, 
 
\begin{theorem}
\label{thm1}
If $\mu_{D10} = \mu_{D20} = \mu_{D0}$,   as $n_{D} \rightarrow \infty$,   
\begin{eqnarray}
\wh{\mu}_{D1} - \mu_{D0} = O_{p}(n_{D}^{-1/8}), \quad  & \wh{\mu}_{D2} - \mu_{D0}=O_{p}(n_{D}^{-1/8}),  \nn \\
 \wh{\rho}_{D} - \rho_{D0}=O_{p}(n_{D}^{-1/4}), \quad  & \wh{\sigma}_{D2}^{2} - \sigma_{D0}^{2} = O_{p}(n_{D}^{-1/4}).  \nn
\end{eqnarray}
\end{theorem}

 \subsection{Combined mixture likelihood estimation} 

  Let $(\mu_{M0}, \rho_{M0},  \mu_{D10}, \mu_{D20}, \rho_{D0}, \sigma_{0}^{2})$ stand for the true values
  and $(\wh{\mu}_{M}, \wh{\rho}_{M},   \wh{\mu}_{D1}, \wh{\mu}_{D2}, \wh{\rho}_{D}, \wh{\sigma}^{2})$ the MLE's from the combined mixture likelihood
  (\ref{eq:combined}). 
 We develop the asymptotics under the condition $n_{M} / n \rightarrow \lambda$ with $ 0 < \lambda <1$ where $n=n_{M} + n_{D}$.  
If $\mu_{D10} \neq \mu_{D20}$,  as $n \rightarrow \infty$, the MLE's are asymptotically normally distributed with root-$n$ consistency, following the classic theories on MLE \citep{serfling2000}. 
Things are complicated if $\mu_{D10} = \mu_{D20}$.

\begin{theorem}
\label{thm2}
 If $\mu_{D10} = \mu_{D20} = \mu_{D0}$, as $n \rightarrow \infty$,  
\begin{eqnarray*}
&\wh{\mu}_{M} - \mu_{M0} = O_{p}(n^{-1/2}),~~\wh{\mu}_{D1} - \mu_{D0} = O_{p}(n^{-1/4}),~~ \wh{\mu}_{D2} - \mu_{D0} = O_{p}(n^{-1/4}),\\
 &\wh{\rho}_{M} - \rho_{M0}=O_{p}(n^{-1/2}),~~\wh{\rho}_{D} - \rho_{D0}=O_{p}(n^{-1/2}),~~ \wh{\sigma}^{2} - \sigma_{0}^{2} = O_{p}(n^{-1/2}). 
 \end{eqnarray*}
\end{theorem}

 This asymptotic property is different from most existing mixture distribution estimations where MLE's are all of convergence rate lower than $n^{-1/2}$.  In Theorem \ref{thm2}, the MLE's, except $\wh{\mu}_{D1}$ and $\wh{\mu}_{D2}$, attain convergence rate of $n^{-1/2}$. 
A heuristic explanation for the improved efficiency is as follows.  First, $l_{n}^{M}(\mu_{M},   \rho_{M}, \sigma^{2})$ is a regular likelihood function that yields root-$n$ consistent estimation of $\mu_{M}$, $\rho_{M}$ and $\sigma^{2}$.    Second, with $\sigma^{2}$ effectively  known in $l_{n}^{D}(\mu_{D1},  \mu_{D2}, \rho_{D}, \sigma^2)$, the combined likelihood enables faster convergence in DZ-related parameters than mixture likelihood over DZ twins only. 

Here is a summary of the methods for correlation estimation. Pearson's correlation coefficient is consistent for $\rho_{M}$ but not for $\rho_{D}$ if a trait is heterogeneous. Estimators based on mixture models --- both the separate mixture and the combined mixture --- are root-$n$ consistent for $\rho_{M}$, and root-$n$ consistent for $\rho_{D}$ if $\mu_{D1} \neq \mu_{D2}$. If in fact $\mu_{D1} = \mu_{D2}$,  $\wh{\rho}_{D}$ from the combined mixture maintains the convergence rate of $n^{-1/2}$ but that from the separate mixture has a convergence rate of $n^{-1/4}$.

The property in Theorem \ref{thm2} is critical to twin study. In practice, we rarely know whether a trait is homogeneous or heterogeneous, which itself is to be investigated, see section \ref{sec:test}. The convergence rate of $\wh{\rho}_{D}$ under the separate mixture is thus uncertain.  Under the combined mixture, both $\wh{\rho}_{M}$ and  $\wh{\rho}_{D}$ are root-$n$ consistent whether a trait is homogeneous or heterogeneous, thus $\wh{\delta} =$ $\wh{\rho}_{M} - \wh{\rho}_{D}$ is always root-$n$ consistent. Numerical studies further indicate that  $\wh{\delta}$ is close to being symmetrically distributed around $\delta$. Standard statistical approaches, such as bootstrap, are appropriate for statistical inference on $\delta$.

\subsection{Homogeneity test} \label{sec:test}

It is sometimes of interest whether a trait is homogeneous in DZ twins. That is to test the hypothesis 

\begin{equation}
H_{0}: \mu_{D1} = \mu_{D2} {\mbox { versus }} H_{a}: \mu_{D1} \neq \mu_{D2}. \nn
\end{equation}

Homogeneity testing under mixture distribution is difficult due to the intrinsic complexity associated with mixture distribution \citep{Dacunha1999, chen-chen2001,   LiuShao2003, Charnigo2004,   Li-etal2015}.  In twin studies with both MZ and DZ twins, we can test homogeneity based on the combined likelihood (\ref{eq:combined}).  Following Theorem \ref{thm1} and Theorem \ref{thm2}, this homogeneity test should outperform the test based on (\ref{eq:mixture}) over DZ twins only.

Recall that $(\wh{\mu}_{M}, \wh{\rho}_{M}, \wh{\mu}_{D1}, \wh{\mu}_{D2},   \wh{\rho}_{D}, \wh{\sigma}^{2})$ stand for the MLE's  under (\ref{eq:combined}). Let $(\wh{\mu}_{M0}, \wh{\rho}_{M0}, \wh{\mu}_{D0},  \wh{\rho}_{D0}, \wh{\sigma}_{0}^{2})$  be the MLE's under the null hypothesis with $\mu_{D10} = \mu_{D20} = \mu_{D0}$. The likelihood ratio test statistic  is
\[
 R^{C}_{n} = 2\{ l^{C}_{n}(\wh{\mu}_{M}, \wh{\rho}_{M}, \wh{\mu}_{D1}, \wh{\mu}_{D2}, \wh{\rho}_{D}, \wh{\sigma}^{2}) - l^{C}_{n}(\wh{\mu}_{M0}, \wh{\rho}_{M0}, \wh{\mu}_{D0}, \wh{\mu}_{D0}, \wh{\rho}_{D0}, \wh{\sigma}_{0}^{2}) \}.
 \]

\begin{theorem}
\label{thm3}
 Under H$_{0}$, as $n \rightarrow \infty$,   $R^{C}_{n}$ converges in distribution to $0.5 \chi_{0}^{2} + 0.5\chi_{1}^{2}$, where $\chi_{0}^{2}$ is  the distribution with point density at zero and $\chi_{1}^{2}$ the chi-squared distribution with one degree of freedom. 
\end{theorem}

 Since convergence rates of $\wh{\mu}_{D1}$ and $\wh{\mu}_{D2}$ are lower than root-$n$, they may affect the approximation accuracy of the limiting distribution in Theorem \ref{thm3} when $n$ is small.  We adjust the limiting distribution by $(1- a_{n})\chi_{0}^{2} + a_{n} \chi_{1}^{2}$ with $a_{n} = E(R_{n}^{C})$. This ensures the adjusted limiting distribution have the same first moment as $R_{n}^{C}$ in the spirit of the Bartlett correction \citep{Bartlett1937}.  
Derivation of the analytical form of $a_{n}$ is challenging and we use a computer experiment method to find an empirical formula for $a_n$ as shown in section \ref{sec:simulation}.  

The proofs of all theorems are provided in the Supplementary document.

 \section{Numerical studies} \label{sec:simulation}

In the following simulations,  $\{ (Y_{i1}^{M}, Y_{i2}^{M}): i=1, \cdots, n_{M}\}$ and $\{ (Y_{j1}^{D}, Y_{j2}^{D}): j=1, \cdots, n_{D}\}$ are both generated  from mixture distribution (\ref{eq:mixture}).  Without loss of generality,  we let $\mu_{M}=0$, $\rho_{M}=0.9$, $\mu_{D1}=0$ and $\sigma^2=1$.  We investigate the numerical performance of the maximum likelihood estimators from two mixture models: the separate mixture and the combined mixture. We examine the estimation of the correlation coefficient $\rho_{D}$, the estimation and inference of $\delta=\rho_{M} - \rho_{D}$ which reflects the collective genetic impact and the properties of the likelihood ratio test statistic $R_{n}^{C}$.

 \begin{table}[!ht]
   \caption{Estimation of $\rho_{D}$: estimates followed by standard deviations, from the combined mixture (Combined) and the separate mixture (Mixture)  treating $\mu_{D1}$ and $\mu_{D2}$ as distinct or equal (marked by *),  and Pearson's correlation coefficients
   \label{tab:sim1}}
    \vspace{0.1in}
\centering
\begin{tabular}{llllllllll} \hline
  & & \multicolumn{8}{c}{$\rho_{D}=0.3$} \\
   $(n_M,n_D)$	&  Estimator		& \multicolumn{2}{c}{$\mu_{D2}=1$} & & \multicolumn{2}{c}{$\mu_{D2}=0.5$} & & \multicolumn{2}{c}{$\mu_{D2}=0$}  \\
	\cline{3-4} \cline{6-7} \cline{9-10}
(100,100) 	& Combined &0.290 & 0.148 & &0.335 & 0.155 & &0.386 & 0.136 \\
	& Combined* & &  & & &  & &0.292 & 0.087 \\
	 & Mixture &0.286 & 0.154 & &0.351 & 0.153 & &0.412 & 0.134 \\
	 & Mixture* & &  & & &  & &0.292 & 0.087 \\
	 & Pearson &0.045 & 0.093 & &0.222 & 0.097 & &0.298 & 0.087 \\
(400,400) & Combined &0.299 & 0.076 & &0.309 & 0.090 & &0.351 & 0.080 \\
	 & Combined* & &  & & &  & &0.298 & 0.044 \\
	 & Mixture  &0.294 & 0.089 & &0.322 & 0.097 & &0.378 & 0.094 \\
	 & Mixture* & &  & & &  & &0.298 & 0.045 \\
	 & Pearson &0.040 & 0.047 & &0.224 & 0.047 & &0.300 & 0.045 
\end{tabular}

\vskip 0.2cm

\begin{tabular}{llllllllll} 
 & & \multicolumn{8}{c}{$\rho_{D}=0.5$} \\
   	   &  	& \multicolumn{2}{c}{$\mu_{D2}=1$} & & \multicolumn{2}{c}{$\mu_{D2}=0.5$} & & \multicolumn{2}{c}{$\mu_{D2}=0$}  \\
	\cline{3-4} \cline{6-7} \cline{9-10}
(100,100) & Combined &0.484 & 0.131 & &0.504 & 0.128 & &0.575 & 0.112 \\
 	& Combined* & &  & & &  & &0.495 & 0.071 \\
 	& Mixture &0.485 & 0.127 & &0.514 & 0.125 & &0.581 & 0.113 \\
 	& Mixture* & &  & & &  & &0.495 & 0.074 \\
 	& Pearson &0.202 & 0.092 & &0.406 & 0.081 & &0.501 & 0.074 \\
(400,400) & Combined &0.498 & 0.060 & &0.499 & 0.078 & &0.545 & 0.069 \\
 	& Combined* & &  & & &  & &0.500 & 0.037 \\
 	& Mixture &0.498 & 0.061 & &0.502 & 0.086 & &0.558 & 0.076 \\
 	& Mixture* & &  & & &  & &0.499 & 0.039 \\
 	& Pearson &0.201 & 0.046 & &0.414 & 0.044 & &0.501 & 0.039 
\end{tabular}

\vskip 0.2cm

\begin{tabular}{llllllllll} 
 &  & \multicolumn{8}{c}{$\rho_{D}=0.8$} \\
  	&  		& \multicolumn{2}{c}{$\mu_{D2}=1$} & & \multicolumn{2}{c}{$\mu_{D2}=0.5$} & & \multicolumn{2}{c}{$\mu_{D2}=0$}  \\
	\cline{3-4} \cline{6-7} \cline{9-10}
(100,100) 	& Combined &0.795 & 0.040 & &0.792 & 0.069 & &0.839 & 0.049 \\
 	& Combined* & &  & & &  & &0.797 & 0.035 \\
 	& Mixture &0.793 & 0.041 & &0.794 & 0.066 & &0.839 & 0.050 \\
 	& Mixture* & &  & & &  & &0.796 & 0.038 \\
 	& Pearson &0.435 & 0.069 & &0.694 & 0.052 & &0.799 & 0.038 \\
(400,400)	& Combined &0.800 & 0.020 & &0.789 & 0.037 & &0.825 & 0.029 \\
 	& Combined* & &  & & &  & &0.797 & 0.017 \\
 	& Mixture &0.799 & 0.021 & &0.792 & 0.039 & &0.826 & 0.035 \\
 	& Mixture* & &  & & &  & &0.796 & 0.019 \\
 	& Pearson &0.438 & 0.034 & &0.693 & 0.025 & &0.797 & 0.018 \\ 
\hline
\end{tabular}  
\end{table}

Table \ref{tab:sim1} presents the estimation of $\rho_{D}$ at different values of $\rho_{D}$ and $\mu_{D2}$.   We observe that Pearson's correlation coefficient underestimates $\rho_{D}$ except in the case $\mu_{D1}=\mu_{D2}$.  This observation agrees with (\ref{eq:Pearson}): the more $\mu_{D1}$ differs from $\mu_{D2}$, the more bias.  In all cases,  both the separate and the combined mixture likelihood estimators are consistent. When $\mu_{D1}$ and $\mu_{D2}$ are very different, both MLE's are root-$n$ consistent.  As sample size increases from $100$ to $400$,  the standard deviations all reduce by a half.  The combined likelihood estimates have smaller variation than the separate likelihood estimates.   When $\mu_{D1} = \mu_{D2}$,  the complexity of mixture distribution estimation shows up. We see that $\wh{\rho}_{D}$ from the combined likelihood maintains root-$n$ consistency. However,   $\wh{\rho}_{D}$ from the separate likelihood is not root-$n$ consistent but has a convergence rate around $n^{-1/4}$: the standard deviation at $n_M=n_D=400$ is roughly $0.71$ that at $n_M=n_D=100$.  In this case, we additionally estimate from mixture models restricting $\mu_{D1} = \mu_{D2}$.  We see that both mixture likelihood estimators produce estimates close to Pearson's correlation coefficients, indicating that, for a trait with homogeneity known, the proposed estimators are nearly equivalent to Pearson's with comparable efficiency.   


 \begin{table}[!ht]
  \caption{Estimation of $\delta$: the estimates, the $95\%$ bootstrap confidence intervals (CI), and the coverage probabilities ($\%$) from the combined mixture (Combined), the separate mixture (Mixture), and Pearson's correlation coefficients
  \label{tab:sim2} }
   \vspace{0.1in}
 \begin{tabular}{lllclllcl}  \hline
 $\delta$ 	&		&  Estimate	& CI	& $\%$    					&&  Estimate	& CI	&   $\%$   \\
  		& 		& \multicolumn{3}{c}{$n_M=n_D=100$, $\mu_{D2}=1$}	&& \multicolumn{3}{c}{$n_M=n_D=400$, $\mu_{D2}=1$} \\
				\cline{3-5} \cline{7-9} 
 0.10 	& Combined &0.100& (0.033, 0.201) & 96  &&0.101 & (0.066, 0.141) & 94 \\
 	& Mixture & 0.098 & (0.016, 0.203) & 94    &&0.105 & (0.064, 0.152) & 92  \\
	 & Pearson &0.447 & ( 0.316 , 0.594 ) & 0  &&0.463 & ( 0.395 , 0.535 ) & 0  \\
 0.35  & Combined &0.349 & (0.196, 0.627)& 92  &&0.351 & (0.262 ,0.498) & 94 \\
 	& Mixture &0.370 & (0.200, 0.606) & 96  &&0.357 & (0.266, 0.485) & 95 \\
	 & Pearson &0.666 & ( 0.499 , 0.845 ) & 3  &&0.67 & ( 0.584 , 0.758 ) & 0 \\
0.60 	& Combined &0.627 & (0.355, 0.935) & 91  &&0.595 & (0.458, 0.770) & 93 \\
	& Mixture &0.601 & (0.355, 0.848) & 88  &&0.595 & (0.458, 0.778) & 95 \\
	 & Pearson &0.858 & ( 0.674 , 1.041 ) & 22  &&0.861 & ( 0.767 , 0.953 ) & 0 \\
  	   & 		& \multicolumn{3}{c}{$n_M=n_D=100$, $\mu_{D2}=0.5$}	&& \multicolumn{3}{c}{$n_M=n_D=400$, $\mu_{D2}=0.5$} \\
			\cline{3-5} \cline{7-9}
0.10 	& Combined &0.093 & (0.003, 0.226) & 91  &&0.107 & (0.048, 0.199) & 94 \\
	& Mixture &0.103 & (-0.002, 0.228) & 95  &&0.102 & (0.042, 0.186) & 92 \\
	 & Pearson &0.206 & ( 0.107 , 0.32 ) & 49  &&0.205 & ( 0.155 , 0.257 ) & 0 \\
 0.35  & Combined &0.341 & (0.145, 0.553) & 95  &&0.348 & (0.211, 0.492) & 96 \\
 	& Mixture &0.345 & (0.145, 0.535) & 95  && 0.350 & (0.213, 0.465) & 96 \\
	 & Pearson &0.453 & ( 0.3 , 0.62 ) & 77  &&0.433 & ( 0.358 , 0.512 ) & 49 \\
0.6  	& Combined &0.564 & (0.301, 0.816) & 92  &&0.591 & (0.418, 0.734) & 94  \\
	& Mixture &0.543 & (0.292, 0.755) & 82  &&0.584 & (0.402, 0.730) & 95 \\
	 & Pearson &0.672 & ( 0.492 , 0.856 ) & 88  &&0.684 & ( 0.591 , 0.777 ) & 60 \\
     & 		& \multicolumn{3}{c}{$n_M=n_D=100$, $\mu_{D2}=0$}	&& \multicolumn{3}{c}{$n_M=n_D=400$, $\mu_{D2}=0$} \\
			\cline{3-5} \cline{7-9}
0.10  & Combined &0.107 & (0.050, 0.175) & 90  &&0.100 & (0.071, 0.133) & 97 \\
	& Mixture & 0.062 & (-0.034, 0.152) & 75  &&0.071 & (0.003, 0.121) & 79  \\
	 & Pearson &0.099 & ( 0.021 , 0.184 ) & 96  &&0.101 & ( 0.062 , 0.142 ) & 96 \\
0.35  & Combined &0.348 & (0.237, 0.482) & 91  &&0.347 & (0.289, 0.411) & 93 \\
	& Mixture &0.281 & (0.098, 0.445) & 86  &&0.283 & (0.166, 0.382) & 77 \\
	 & Pearson &0.349 & ( 0.217 , 0.496 ) & 93 &&0.347 & ( 0.281 , 0.419 ) & 96 \\
0.60 	& Combined &0.595 & (0.435, 0.778) & 95  &&0.600 & (0.518, 0.688) & 94 \\
	& Mixture &0.467 & (0.224, 0.687) & 75  &&0.515 & (0.351, 0.645) & 73 \\
	 & Pearson &0.602 & ( 0.431 , 0.782 ) & 95  &&0.601 & ( 0.513 , 0.691 ) & 95 \\
 \hline
\end{tabular}
\end{table}

Table \ref{tab:sim2} presents the estimation of $\delta = $ $\rho_{M} - \rho_{D}$ at different values of $\delta$ and $\mu_{D2}$.  When $\mu_{D1} \neq \mu_{D2}$, Pearson's correlation coefficient overestimates $\delta$.  This observation is consistent with (\ref{eq:Pearson}) as Pearson's correlation coefficient is consistent for $\rho_{M}$ but underestimates $\rho_{D}$.  In all cases, including when $\mu_{D1} = \mu_{D2}$, the combined likelihood estimator is root-n consistent for $\delta$. It is of negligible bias with bootstrap confidence interval having coverage probability around $0.95$.  As sample size increases from $100$ to $400$, the width of confidence intervals reduces about a half. The distribution of $\wh{\delta}$ is further investigated, indicating that the combined likelihood estimator of $\delta$ is close to being symmetrically distributed in all cases, with results shown in the supplementary document. The separate likelihood estimator has similar performance when $\mu_{D1} \neq \mu_{D2}$ but not as good when $\mu_{D1} = \mu_{D2}$.  In the latter case, the separate likelihood estimator of $\rho_{M}$ is root-n consistent but that of $\rho_{D}$ is not, we thus see substantial bias in $\wh{\delta}$ and poor coverage probability in bootstrap confidence intervals at moderate sample size.

We then evaluate the limiting distribution in Theorem \ref{thm3} and the adjusted limiting distribution $a_{n} \chi_{0}^{2} + (1-a_{n}) \chi_{1}^{2}$.  In the following, we adopt a computer experiment method to derive an empirical formula of $a_{n}$ as function of $n$.  Under $H_{0}$, both $(Y_{i1}^{M}, Y_{i2}^{M})$ and $(Y_{j1}^{D}, Y_{j2}^{D})$ are bivariate normally distributed. Since $R_n^C$ is invariant to location and orthogonal transformations, 
we consider the null model to be the standard bivariate normal distribution for  both $(Y_{i1}^{M}, Y_{i2}^{M})$ and $(Y_{j1}^{D}, Y_{j2}^{D})$. 

\begin{table}
 \caption{$\hat a_n$  over $N = 10000$ sets of random samples} \label{an}
 \vspace{0.1in}
\begin{center}
\begin{tabular}{c|cccccccc}
\hline
$n_M\setminus n_D$ &50&100&150&200&250&300&350&400\\
\hline 
50&0.61&0.62&0.63&0.63&0.63&0.64&0.63&0.64\\
100&0.56&0.57&0.58&0.58&0.59&0.58&0.60&0.58\\
150&0.55&0.54&0.55&0.55&0.55&0.56&0.56&0.57\\
200&0.53&0.53&0.53&0.54&0.55&0.54&0.54&0.55\\
250&0.52&0.52&0.52&0.53&0.53&0.53&0.53&0.54\\
300&0.52&0.51&0.52&0.51&0.52&0.52&0.53&0.53\\
350&0.51&0.50&0.51&0.51&0.52&0.51&0.52&0.52\\
400&0.50&0.49&0.49&0.51&0.52&0.51&0.52&0.52\\
\hline
\end{tabular}
\end{center}
\end{table}

\begin{enumerate}
\item For every $n_M,n_D\in \{50, 100, 150,  200, 250, 300, 350, 400\}$, we do the following:
\begin{enumerate}
\item Generate $N$ independent sets of random samples; each has $\{ (Y_{i1}^{M}, Y_{i2}^{M}): i=1, \cdots, n_{M}\}$ and  $\{ (Y_{j1}^{D}, Y_{j2}^{D}): j=1, \cdots, n_{D}\}$ from standard bivariate normal distribution.
\item Calculate the likelihood ratio test statistic $R_n^C$ for each set of the random samples in (a).
\item Calculate the average of the $R_n^C$'s from the $N$ data sets, which is denoted by $\hat a_n$.
\end{enumerate}
The resulting $\hat a_n$ values based on $N=10000$ are summarized in Table \ref{an}.

\item After some exploratory analysis, we find that $n_D$ has little effect on $\hat a_n$ and 
$1/n_M$ is the only important covariate for modeling $\hat a_n$.  
Because of that, we fit a regression model of $\hat a_n$ on $n$ as follows:  
\[
 \hat a_n = 0.5+a/n_M+\epsilon_n.
\]
With the $\hat a_n$ values in Table \ref{an}, the estimated $a$ is 6.828. 
Hence we recommend the following empirical formula for $a_n$: 
$$
a_n=0.5+ 6.828/n_M. 
$$
\end{enumerate}

Again as $R_{n}^{C}$ is invariant to location and orthogonal  transformations, the above formula applies to general combined mixture distribution.

\begin{figure}[!ht]
 \centering
 \includegraphics[width=5in,height=6in, keepaspectratio]{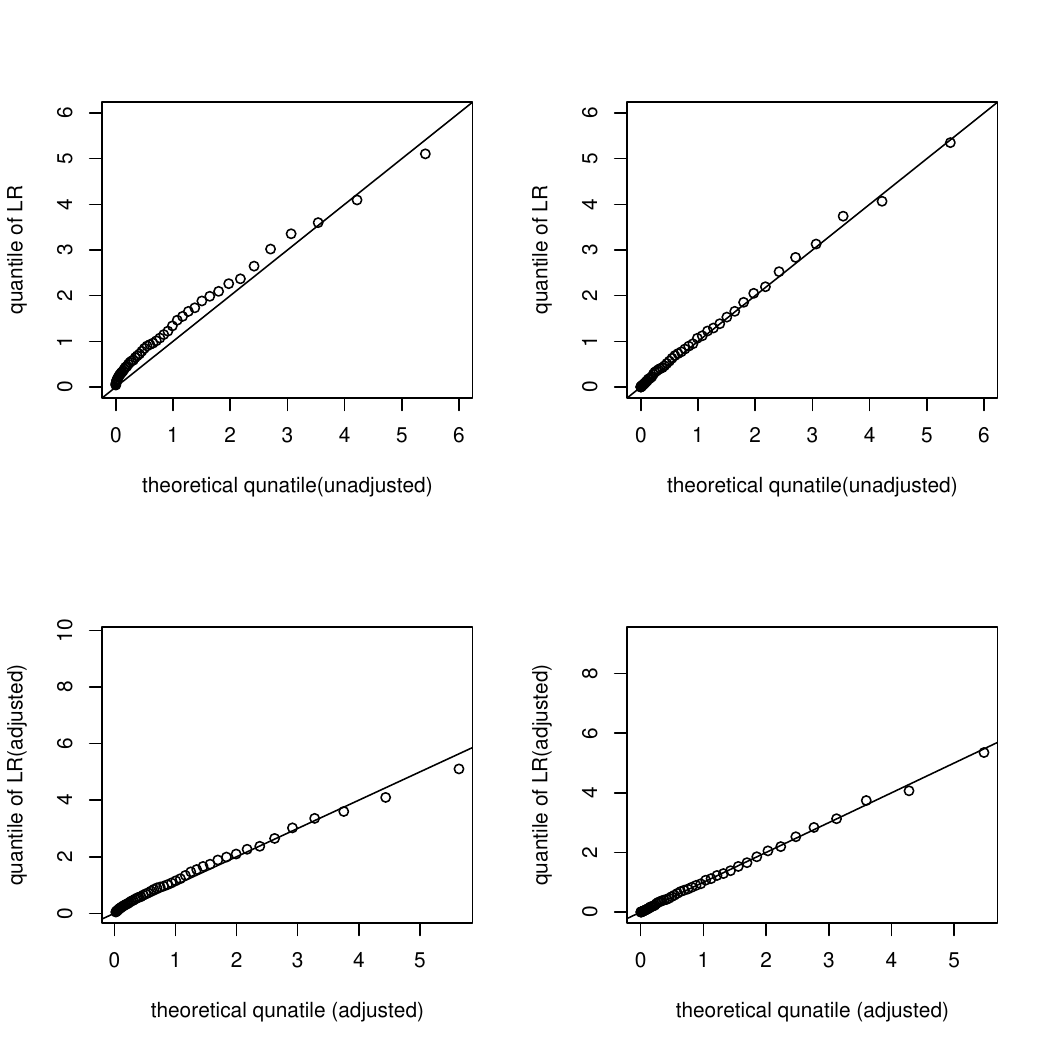}
  \caption{Quantile-quantile plot of $R_{n}^{C}$ under H$_{0}$: the unadjusted limiting distribution on the top, the adjusted limiting distribution on the bottom, $n_M=n_D=100$ on the left and $400$ on the right,  at $\rho_{M} = 0.9$ and $\rho_{D} = 0.3$.   }
  \label{fig:qqplot}
 \end{figure}

Figure \ref{fig:qqplot} presents the quantile-quantile plots of $R_{n}^{C}$ under $H_{0}$ against the limiting distribution $0.5\chi_{0}^{2} + 0.5 \chi_{1}^{2}$ and the adjusted limiting distribution $a_{n} \chi_{0}^{2} + (1-a_{n}) \chi_{1}^{2}$.  
Both both limiting distributions fit the null distribution of $R_{n}^{C}$ pretty well.  The adjusted limiting distribution has a slightly better approximation when sample size is small.  
  

Additional numerical study outcomes can be found in the supplementary document. 

\section{Application to twin study on immune traits} \label{sec:application}

In the immune trait study introduced in section \ref{sec:intro}, we are interested in six major immune traits:  the frequencies of total lymphocytes, B cells, total T cells,  CD4 T cells,  CD8 T cells, and CD27 T cells.  Lymphocytes are responsible for adaptive immune responses and mainly consist of B cells and T cells. B cells produce antibodies that recognize foreign substances, but it is the job of T cells to attack them. CD4 T cells help other immune cells to respond to extracellular sources of infection, CD8 T cells kill the infected cells, and CD27 T cells are crucial for the generation of T cell memory.  For a specific trait, the observed are data $(P_{i1}, P_{i2})$, the frequency of a specific immune cell subset in the $i$-th twin pair.  We assess the genetic impact on each trait. 


We apply the combined mixture model to the twin data. According to section \ref{sec:mixture},  the combined mixture model is appropriate for paired data under two assumptions. The first assumption is normality of the underlying ordered pairs $ \{(Y_{i1}^{*}, Y_{i2}^{*}), i=1, \cdots, n\}$. In this study the  observed data, $(P_{i1}, P_{i2})$, are frequencies that range from $0$ to $1$, we need to first transform the data to be normally distributed. 
 We  find the  transformation $Y_{ij} = \Phi^{-1}(P_{ij})$, with $\Phi$ the cumulative distribution function of the standard normal distribution, works well.   For MZ twin data $(Y_{i1}, Y_{i2})$, the order within a pair does not matter and it has the same distribution as the underlying ordered pair $(Y_{i1}^{*}, Y_{i2}^{*})$.  We thus evaluate the appropriateness of the transformation over MZ twins. 
  If $\{ (Y_{i1}, Y_{i2}), i=1, \cdots, n\}$ are normally distributed,   $Y_{i1}$ and $Y_{i2}$ are of identical normal distribution.  We check on normality as follows. For each trait, 1) randomly pick one measure $Y_{ii_{r}}$, with $i_{r}=1$ or $2$,  from each twin pair; 2) examine the normality of $\{Y_{ii_{r}}, i=1, \cdots, n \}$ via Shapiro-wilk test; 3) repeat the two steps multiple times and compute the mean p value in step 2).  We find the mean p values in the range of $0.48$ to $0.63$ for all traits, indicating the transformed data are close to being normally distributed. Since the transformation normalizes the MZ twin data,  it is reasonable to believe the same transformation normalizes the DZ twin data.  
 
 The other assumption is that MZ and DZ twin data share the same variance, or $\sigma^{2}_{M} = \sigma^{2}_{D}$. To check on this, we estimate $\sigma^{2}_{M}$ and $\sigma^{2}_{D}$ from the separate mixture model (\ref{eq:mixture}) that is free of the equal variance assumption.  The variance estimates, $\wh{\sigma}^{2}_{M}$  and $ \wh{\sigma}^{2}_{D}$, are quite close in all traits, indicating no violation of the assumption.  A more rigorous approach for examining this assumption  would be testing the null hypothesis $\sigma^{2}_{M} = \sigma^{2}_{D}$ via a likelihood ratio test with the log-likelihood ratio $2[ \{ l_{n}^{M}(\wh{\mu}_{M}, \wh{\rho}_{M}, \wh{\sigma}_{M}^{2})  + l_{n}^{M}(\wh{\mu}_{D1}, \wh{\mu}_{D2}, \wh{\rho}_{D}, \wh{\sigma}_{D}^{2}) \} - l_{n}^{C}(\wh{\mu}_{M}, \wh{\rho}_{M}, \wh{\mu}_{D1}, \wh{\mu}_{D2}, \wh{\sigma}^{2}) ]$, where the first term is the log-likelihood under the separate mixture model and the second term under the combined mixture model.  However, the limiting distribution of the test statistic is difficult to derive and beyond the scope of this work. 
    

\begin{table}[!ht]
\centering
\caption{Outcomes over six immune traits: estimates of $\rho_{M}$, $\rho_{D}$, $\delta$ and the $95\%$ bootstrap confidence interval (CI) from the combined mixture  and Pearson's correlation coefficients.
\label{tab:example}}
\vspace{0.1in}
\begin{tabular}{lllllc}
  \hline
  Index & Trait  & Estimator & $\wh{\rho}_{M}$ & $\wh{\rho}_{D}$ & $\wh{\delta}$ (CI)  \\ 
  \hline
1  & $\%$ L  		& Combined 	& 0.56 & 0.44 & 0.12 (0.02, 0.25)  \\ 
     &				& Pearson   	& 0.56 & 0.45 & 0.11 (0.01, 0.24) 	   \\ 
2 & $\%$ L: B 		& Combined & 0.54 & 0.38 & 0.16 (0.05, 0.38)   \\ 
   &				& Pearson    & 0.53 & 0.11 & 0.42 (0.29, 0.56)    \\ 				
3 & $\%$ L: T 		& Combined & 0.57 & 0.28 & 0.29 (0.16, 0.41)   \\ 
   &				& Pearson & 0.56 & 0.29 & 0.27 (0.13, 0.42)     \\ 				
 4 & $\%$ L: CD27	& Combined & 0.75 & 0.40 & 0.35 (0.15, 0.53)  \\ 
    &				& Pearson & 0.74 & 0.26 & 0.48 (0.37, 0.64)     \\ 
5  & $\%$ L: CD8 	& Combined 	& 0.71 & 0.30 & 0.41 (0.26, 0.60) 	 \\ 
    &				& Pearson   & 0.72 & 0.13 & 0.60  (0.50, 0.72)      \\ 					
6 & $\%$ L: CD4 	& Combined & 0.77 & 0.25 & 0.52 (0.39, 0.69)   \\ 
   &				& Pearson & 0.74 & 0.26 & 0.48 (0.34, 0.65)     \\ 
\hline
\end{tabular}
\end{table}


Table \ref{tab:example} presents the estimation of genetic impact on the six immune traits:  the frequencies of total lymphocytes ($\%$ L) and among the lymphocytes, the frequencies of B cells ($\%$ L: B), T cells ($\%$ L: T), CD4 T cells ($\%$ L: CD4), CD8 T cells ($\%$ L: CD8), and CD27 T cells ($\%$ L: CD27).     We observe that  estimates of $\delta$ are all positive, indicating genetic contribution to all these traits.   The genetic impact on  the three T cell subsets --- CD27, CD4 and CD8  --- are higher than the genetic impact on the others, indicating these traits are likely genetically determined. This result resonates with the findings of other studies that, though variability in human immune system is largely driven by non-genetic influences such as age and season, variations in T cell subsets are relatively more affected by genetic factors \citep{Brodin2015, Gamboa2016}.

\begin{figure}[!ht]
 \centering
 \includegraphics[width=5.5in,height=4in]{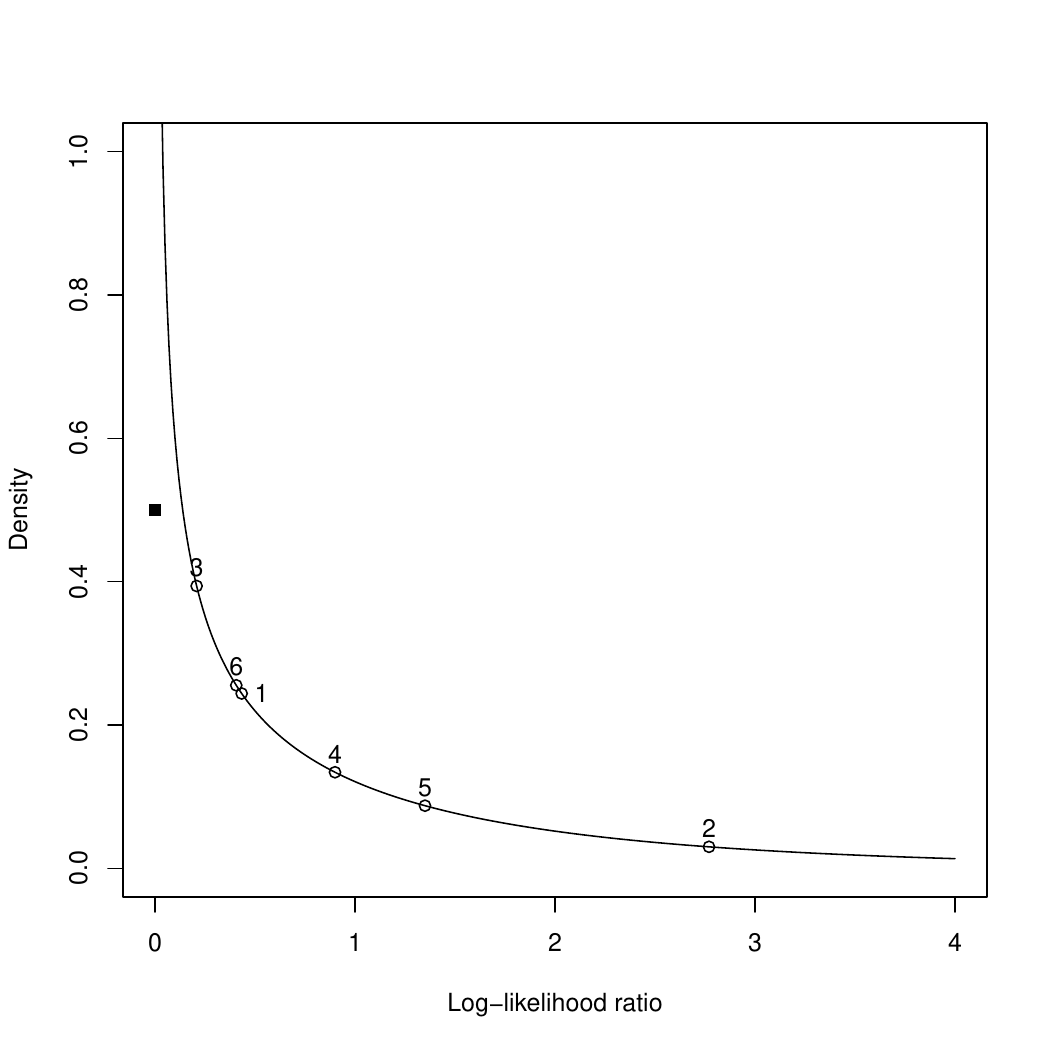}
  \vskip 6mm
  \caption{Homogeneity test results:  the observed log-likelihood ratios for the six immune traits and the density curve of the limiting distribution.}
  \label{fig:example}
 \end{figure}

 Figure \ref{fig:example} shows homogeneity test results: the observed log-likelihood ratios along with the density curve of the limiting distribution of   the  log-likelihood ratio under the null.  According to Theorem \ref{thm3}, the limiting distribution is a mixture chi-squared distribution with a point density at zero.  The more the log-likelihood ratio lies to the right tail of the limiting distribution, the stronger is the evidence of heterogeneity.   Figure \ref{fig:example} indicates lack of homogeneity in some immune traits, such as the frequency of B cells, CD4 T cells and CD8 T cells, though of only marginal statistical significance.   This also explains why Pearson's correlation coefficients in these immune traits differ from the mixture model estimates in Table \ref{tab:example}: for a heterogeneous trait, Pearson's  correlation underestimates $\rho_{D}$, the correlation coefficient in DZ twins,  and consequently overestimates the genetic impact $\delta=\rho_{M} - \rho_{D}$. 
 
 The data that support the findings of this study are available on request from the corresponding author. The data are not publicly available due to privacy or ethical restrictions.

 



\bibliographystyle{biometrika}
\bibliography{ref}  

\end{document}